\documentclass[aps,prd,showpacs,showkeys,amsmath,amssymb,
twocolumn,
floatfix,
superscriptaddress
]{revtex4-1}
\usepackage{graphicx}
\usepackage{times}
\usepackage{multirow}
\usepackage{verbatim}
\usepackage{color,soul}
\usepackage{xcolor}
\usepackage{cancel}
\usepackage[normalem]{ulem}

\graphicspath{{figures/}}

\begin{document}


\title{Evolution of the Reactor Antineutrino Flux and Spectrum at Daya Bay}

\newcommand{\ECUST}{\affiliation{Institute of Modern Physics, East China University of Science and Technology, Shanghai}}
\newcommand{\IHEP}{\affiliation{Institute~of~High~Energy~Physics, Beijing}}
\newcommand{\Wisconsin}{\affiliation{University~of~Wisconsin, Madison, Wisconsin 53706, USA}}
\newcommand{\Yale}{\affiliation{Wright~Laboratory and Department~of~Physics, Yale~University, New~Haven, Connecticut 06520, USA}} 
\newcommand{\BNL}{\affiliation{Brookhaven~National~Laboratory, Upton, New York 11973, USA}}
\newcommand{\NTU}{\affiliation{Department of Physics, National~Taiwan~University, Taipei}}
\newcommand{\NUU}{\affiliation{National~United~University, Miao-Li}}
\newcommand{\Dubna}{\affiliation{Joint~Institute~for~Nuclear~Research, Dubna, Moscow~Region}}
\newcommand{\CalTech}{\affiliation{California~Institute~of~Technology, Pasadena, California 91125, USA}}
\newcommand{\CUHK}{\affiliation{Chinese~University~of~Hong~Kong, Hong~Kong}}
\newcommand{\NCTU}{\affiliation{Institute~of~Physics, National~Chiao-Tung~University, Hsinchu}}
\newcommand{\NJU}{\affiliation{Nanjing~University, Nanjing}}
\newcommand{\TsingHua}{\affiliation{Department~of~Engineering~Physics, Tsinghua~University, Beijing}}
\newcommand{\SZU}{\affiliation{Shenzhen~University, Shenzhen}}
\newcommand{\NCEPU}{\affiliation{North~China~Electric~Power~University, Beijing}}
\newcommand{\Siena}{\affiliation{Siena~College, Loudonville, New York  12211, USA}}
\newcommand{\IIT}{\affiliation{Department of Physics, Illinois~Institute~of~Technology, Chicago, Illinois  60616, USA}}
\newcommand{\LBNL}{\affiliation{Lawrence~Berkeley~National~Laboratory, Berkeley, California 94720, USA}}
\newcommand{\UIUC}{\affiliation{Department of Physics, University~of~Illinois~at~Urbana-Champaign, Urbana, Illinois 61801, USA}}
\newcommand{\RPI}{\affiliation{Department~of~Physics, Applied~Physics, and~Astronomy, Rensselaer~Polytechnic~Institute, Troy, New~York  12180, USA}}
\newcommand{\SJTU}{\affiliation{Department of Physics and Astronomy, Shanghai Jiao Tong University, Shanghai Laboratory for Particle Physics and Cosmology, Shanghai}}
\newcommand{\BNU}{\affiliation{Beijing~Normal~University, Beijing}}
\newcommand{\WM}{\affiliation{College~of~William~and~Mary, Williamsburg, Virginia  23187, USA}}
\newcommand{\Princeton}{\affiliation{Joseph Henry Laboratories, Princeton~University, Princeton, New~Jersey 08544, USA}}
\newcommand{\VirginiaTech}{\affiliation{Center for Neutrino Physics, Virginia~Tech, Blacksburg, Virginia  24061, USA}}
\newcommand{\CIAE}{\affiliation{China~Institute~of~Atomic~Energy, Beijing}}
\newcommand{\SDU}{\affiliation{Shandong~University, Jinan}}
\newcommand{\NanKai}{\affiliation{School of Physics, Nankai~University, Tianjin}}
\newcommand{\UC}{\affiliation{Department of Physics, University~of~Cincinnati, Cincinnati, Ohio 45221, USA}}
\newcommand{\DGUT}{\affiliation{Dongguan~University~of~Technology, Dongguan}}
\newcommand{\XJTU}{\affiliation{Department of Nuclear Science and Technology, School of Energy and Power Engineering, Xi'an Jiaotong University, Xi'an}}
\newcommand{\UCB}{\affiliation{Department of Physics, University~of~California, Berkeley, California  94720, USA}}
\newcommand{\HKU}{\affiliation{Department of Physics, The~University~of~Hong~Kong, Pokfulam, Hong~Kong}}
\newcommand{\UH}{\affiliation{Department of Physics, University~of~Houston, Houston, Texas  77204, USA}}
\newcommand{\Charles}{\affiliation{Charles~University, Faculty~of~Mathematics~and~Physics, Prague, Czech~Republic}} 
\newcommand{\USTC}{\affiliation{University~of~Science~and~Technology~of~China, Hefei}}
\newcommand{\TempleUniversity}{\affiliation{Department~of~Physics, College~of~Science~and~Technology, Temple~University, Philadelphia, Pennsylvania  19122, USA}}
\newcommand{\CUC}{\affiliation{Instituto de F\'isica, Pontificia Universidad Cat\'olica de Chile, Santiago, Chile}} 
\newcommand{\CGNPG}{\affiliation{China General Nuclear Power Group}}
\newcommand{\NUDT}{\affiliation{College of Electronic Science and Engineering, National University of Defense Technology, Changsha}} 
\newcommand{\IowaState}{\affiliation{Iowa~State~University, Ames, Iowa  50011, USA}}
\newcommand{\ZSU}{\affiliation{Sun Yat-Sen (Zhongshan) University, Guangzhou}}
\newcommand{\CQU}{\affiliation{Chongqing University, Chongqing}} 
\newcommand{\BCC}{\altaffiliation[Now at ]{Department of Chemistry and Chemical Technology, Bronx Community College, Bronx, New York  10453, USA}} 
\author{F.~P.~An}\ECUST
\author{A.~B.~Balantekin}\Wisconsin
\author{H.~R.~Band}\Yale
\author{M.~Bishai}\BNL
\author{S.~Blyth}\NTU\NUU
\author{D.~Cao}\NJU
\author{G.~F.~Cao}\IHEP
\author{J.~Cao}\IHEP
\author{Y.~L.~Chan}\CUHK
\author{J.~F.~Chang}\IHEP
% ...


\collaboration{The Daya Bay Collaboration}\noaffiliation
\date{\today}

\begin{abstract}
The Daya Bay experiment has observed correlations between reactor core fuel evolution and changes in the reactor antineutrino flux and energy spectrum.  
Four antineutrino detectors in two experimental halls were used to identify 2.2 million inverse beta decays (IBDs) over 1230 days spanning multiple fuel cycles for each of six 2.9 GW$_{\textrm{th}}$ reactor cores at the Daya Bay and Ling Ao nuclear power plants.  
Using detector data spanning effective $^{239}$Pu fission fractions $F_{239}$ from 0.25 to 0.35, Daya Bay measures an average IBD yield, $\bar{\sigma}_f$, of $(5.90 \pm 0.13) \times 10^{-43}$~cm$^2$/fission and a fuel-dependent variation in the IBD yield, $d\sigma_f/dF_{239}$, of $(-1.86 \pm 0.18) \times 10^{-43}$~cm$^2$/fission.  
This observation rejects the hypothesis of a constant antineutrino flux as a function of the $^{239}$Pu fission fraction at 10 standard deviations.  
The variation in IBD yield is found to be energy dependent, rejecting the hypothesis of a constant antineutrino energy spectrum at 5.1 standard deviations.  
While measurements of the evolution in the IBD spectrum show general agreement with predictions from recent reactor models, the measured evolution in total IBD yield disagrees with recent predictions at 3.1$\sigma$.  
This discrepancy indicates that an overall deficit in measured flux with respect to predictions does not result from equal fractional deficits from the primary fission isotopes $^{235}$U, $^{239}$Pu, $^{238}$U, and $^{241}$Pu.
Based on measured IBD yield variations, yields of $(6.17 \pm 0.17)$ and $(4.27 \pm 0.26) \times 10^{-43}$~cm$^2$/fission have been determined for the two dominant fission parent isotopes $^{235}$U and $^{239}$Pu.  
A 7.8\% discrepancy between the observed and predicted $^{235}$U yields suggests that this isotope may be the primary contributor to the reactor antineutrino anomaly.
\end{abstract}

\maketitle


Electron antineutrinos are produced in commercial nuclear reactor cores as neutron-rich  fission fragments of the fission isotopes $^{235}$U, $^{238}$U, $^{239}$Pu, and $^{241}$Pu beta decay successively toward the isotopic line of stability.  
The total electron antineutrino flux produced by a reactor core is the sum of  thousands of individual beta decay branches, each producing its unique antineutrino flux and spectrum.  
Daya Bay has recently reported measurements of this aggregate antineutrino flux and spectrum~\cite{bib:prl_reactor,bib:cpc_reactor}.  
These measurements confirm the observed discrepancy of $\sim$6\% between the measured reactor antineutrino fluxes of past experiments and reactor model predictions~\cite{bib:huber,bib:mueller2011}, also known as the ``reactor antineutrino anomaly''~\cite{bib:mention2011}, and indicate a disagreement between the measured and predicted antineutrino energy spectrum in the energy range of 5-7 MeV. 
Similar results have also been reported by other current reactor experiments~\cite{dc_bump, reno_bump}.  
Existing interpretations for these flux and spectrum discrepancies include  deficiencies in fission beta spectrum conversion inputs and nuclear databases~\cite{bib:hayes,bib:dwyer,bib:hayes,hayes2,sonzongi2} or the existence of sterile neutrinos~\cite{bib:kopp}.  
If correct, these explanations could have implications for future neutrino experiments~\cite{Kayser,palazzoDune} and nuclear applications~\cite{bib:IAEA}.  

One factor taken into account but not yet directly measured in Daya Bay analyses is the effect of fuel evolution on the observed reactor antineutrino spectrum.  
Since fission yields and beta decay branches from each fission parent isotope are not identical, antineutrino fluxes and spectra produced from the various fission isotopes differ~\cite{VogelHayesReview}.  
Thus, when a reactor experiences a change in the percent contribution to fission rates from each fissioning isotope (fission fractions), a measurable change in the reactor antineutrino flux and spectrum may also be produced.  
Previous experiments have demonstrated variations in the total reactor antineutrino flux with fuel evolution~\cite{bowdenmon,rovnomon}, while providing indications that a change in the spectral shape with fuel evolution may be present~\cite{rovnomon}.  
In this Letter, we report the direct observation of a change in the reactor antineutrino flux and spectrum with reactor fuel evolution.  
This result is then used to determine the reactor antineutrino flux produced by $^{235}$U and $^{239}$Pu and to perform new tests of reactor antineutrino models.  

The Daya Bay Reactor Neutrino Experiment studies the flux of electron antineutrinos produced by six 2.9 GW$_{\textrm{th}}$ commercial reactor cores in two near experimental halls (EH1 and EH2) and one far experimental hall (EH3)~\cite{bib:Detector_NIM}.  
EH3 houses four antineutrino detectors (ADs), while EH1 and EH2 each house two. 
Only the data acquired with the four ADs in EH1 and EH2 in a period covering 1230 days from 2011 to 2015 were utilized in this analysis.  
This includes a period of 217 days with only three ADs present in the near halls, before the second AD was installed in EH2.  
EH1 is situated at a distance of about $\sim$360~m from two cores, while EH2 is $\sim$500~m away from the other four.  
Antineutrinos were detected via the inverse beta decay (IBD) reaction, $\bar{\nu}_e + p \rightarrow e^+ + n$.  
An IBD candidate was defined as a time-correlated trigger pair consisting of a prompt $e^+$ candidate with reconstructed  energy $E_p \approx E_{\nu} - 0.8\ {\rm MeV}$ between 0.7 and 12~MeV and a delayed candidate from neutron capture on gadolinium in the target with 6-12~MeV reconstructed energy~\cite{bib:prd_osc}.  
An IBD candidate set was required to be isolated in time from cosmogenic muon activity or any other AD triggers.  
This selection produced a set of about 1,198,000 and 1,025,000 IBD candidates from EH1 and EH2, respectively.  

Accidental time coincidences of uncorrelated triggers, the dominant background in all ADs, contribute a rate of $\sim$1\% the size of the IBD signal.  
To account for the $<$10\% variations in the rate of this background with time, it was calculated and subtracted week by week for each AD.  
The remaining backgrounds, which contribute $\sim$0.5\% of IBD candidates, were subtracted assuming no time variation in shape or normalization.  

The spectrum of reactor antineutrinos with energy $E_{\nu}$ detected by an AD at time $t$ is expected to be 
\begin{equation}\label{equ_ibd_rate}
\frac{d^2N(E_{\nu},t)}{dE_{\nu}dt} = N_p  \sigma(E_{\nu}) \varepsilon \sum_{r=1}^6 \frac{
\!P(E_{\nu},L_r)}{4\pi L_r^2} 
 \frac{d^{2}\phi_r(E_{\nu}, t)}{dE_{\nu}dt}
\end{equation}
where 
$N_p$ is the number of target protons,
$\sigma(E_{\nu})$ is the IBD reaction cross section, 
$\varepsilon$ is the efficiency of detecting IBDs, 
$L_r$ is the distance between the centers of the AD and the $r$-th core,  
and $P(E_{\nu},L_r)$ is the survival probability due to neutrino oscillation from core $r$.  
The sum in $r$ is taken over the six reactor cores present at Daya Bay.  
The term $d^{2}\phi_r(E_{\nu}, t)/dE_{\nu} dt$ is the antineutrino spectrum from the $r$-th reactor core:
\begin{equation}\label{equ_antinu_prod}
\resizebox{\hsize}{!}{$\displaystyle\frac{d^{2}\phi_r(E_{\nu}, t)}{dE_{\nu} dt} = \frac{W_{\mathrm{th},r}(t)}{\overline{E}_{r}(t)}\sum_{i}f_{i,r}(t)s_{i}(E_{\nu})c^{\mathrm{ne}}_{i}(E_{\nu}) + s_\textrm{SNF}(E_{\nu}),$}
\end{equation}
where the index $i$ runs over the four primary fission isotopes ($^{235}$U, $^{238}$U, $^{239}$Pu, and $^{241}$Pu), $W_{\mathrm{th}}(t)$ is the reactor thermal power, 
$f_{i}(t)$ is the fraction of fissions from isotope $i$,
$\overline{E}_{r}(t) = \sum_{i}f_{i,r}(t)e_{i}$ is the core's average energy released per fission due to the average energy release $e_i$ from each fission isotope, and
$s_{i}(E_{\nu})$ is the $\bar{\nu}_{e}$ energy spectrum per fission.  
All other fission isotopes contribute $<$0.3\% to the total antineutrino flux~\cite{bib:cpc_reactor}, and are neglected in this analysis.  
The correction $c^{\mathrm{ne}}_{i}(E_{\nu})$ accounts for reactor nonequilibrium effects of long-lived fission fragments, and $s_\textrm{SNF}(E_{\nu})$ is the contribution from nearby spent nuclear fuel; both of these quantities are treated as time independent, an assumption that has a negligible impact on the analysis.  

The evolution of the antineutrino flux and spectrum was studied as a function of the effective fission fractions $F_i(t)$ viewed by each AD:
\begin{equation}
\label{eq:effFf}
F_{i}(t) = \sum_{r=1}^6\frac{W_{\mathrm{th},r}(t) \bar{p}_r f_{i,r}(t)}{L^{2}_{r} \overline{E}_{r}(t)}
\bigg/
\sum_{r=1}^6\frac{W_{\mathrm{th},r}(t)\overline{p}_r}{L^{2}_{r}\overline{E}_{r}(t)}.
\end{equation}
The mean survival probability $\bar{p}_r$, calculated by integrating the flux- and cross-section-weighted oscillation survival probability of antineutrinos from core $r$ over $E_{\nu}$, is treated as time independent.  
The four effective fission fractions $F_{235}$, $F_{238}$, $F_{239}$, and $F_{241}$, corresponding to the $^{235}$U, $^{238}$U, $^{239}$Pu, and $^{241}$Pu isotopes respectively, sum to unity at all times for any AD.  
The definition in Eq.~\ref{eq:effFf} allows the expression of the measured IBD yield per nuclear fission $\sigma_f$ as a simple sum of IBD yields from the individual isotopes, $\sigma_f = \sum_i F_i \sigma_i$.  
Weekly effective fission fraction values for each detector were produced using thermal power and fission fraction data for each core, which were provided by the power plant and validated by the Collaboration using the APOLLO2 reactor modeling code~\cite{bib:cpc_reactor}.  
The baselines and the mean survival probabilities used are the same as in Ref.~\cite{bib:prd_osc}, while $e_i$ values were taken from Ref.~\cite{bib:fr_ma}.

\begin{figure}[htb!pb]
\centering
\includegraphics[width=0.9\linewidth]{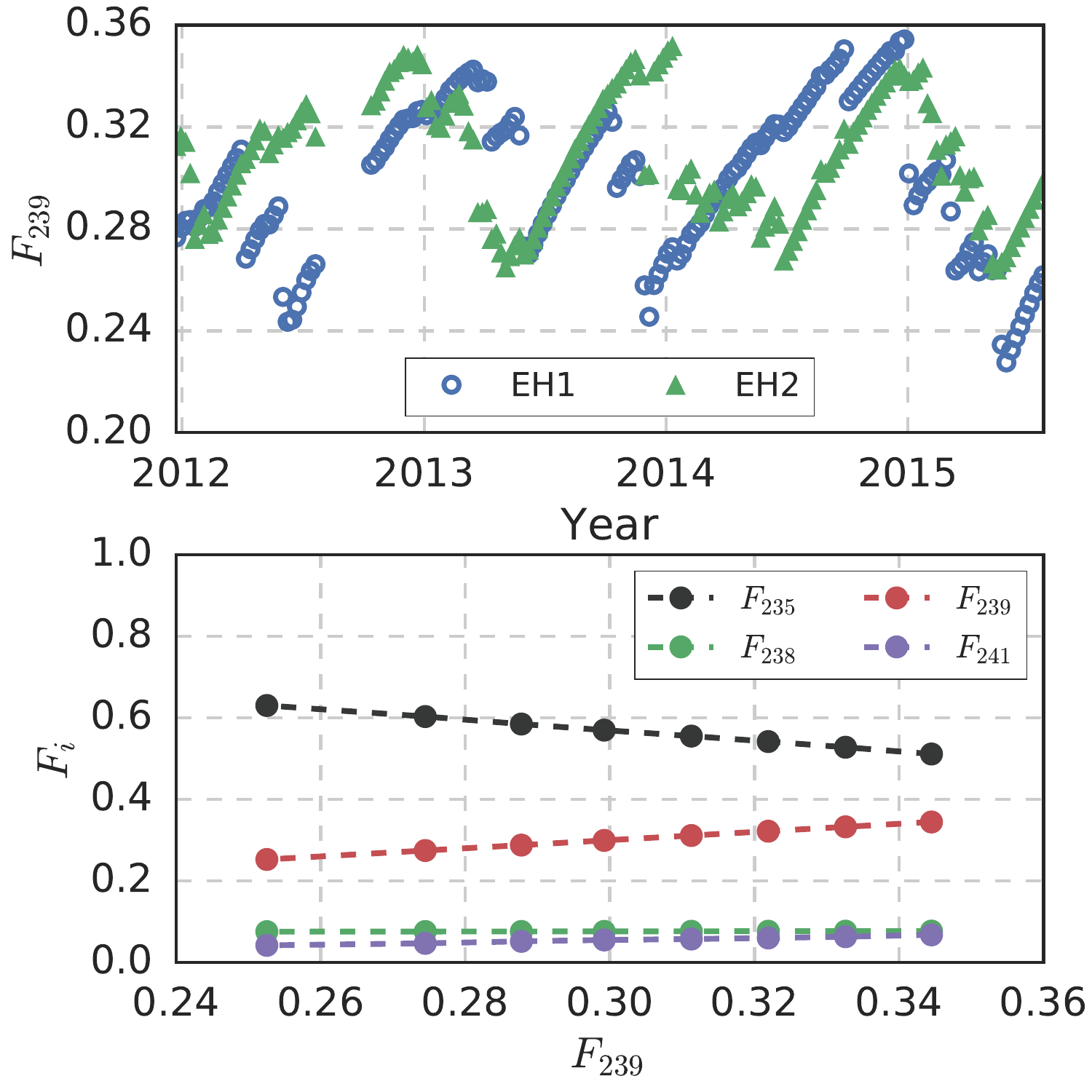}
\caption{Top: Weekly effective $^{239}$Pu fission fractions $F_{239}$ (defined in Eq.~\ref{eq:effFf}) for the EH1 and EH2 ADs based on input reactor data.  
Bottom: Effective fission fractions for the primary fission isotopes versus $F_{239}$.  Each data point represents an average over periods of similar $F_{239}$ from the top panel.} 
\label{fig:Fracs}
\end{figure}

Throughout the Letter, changes in the IBD yield and spectrum per fission will be represented as a function of the effective fission fraction $F_{239}$, which increases as nearby reactors' fuel cycles progress.    
At the beginning of each core's fuel cycle, when 1/3 (1/4) of the fuel rods in the Daya Bay (Ling Ao) cores are fresh, $^{239}$Pu fission fractions $f_{239}$ are $\sim$15\%.  
This fraction then rises to $\sim$40\% by the end of the cycle.
Effective $^{239}$Pu fission fractions $F_{239}$ are shown for the EH1 and EH2 ADs in Fig.~\ref{fig:Fracs}.  
The $F_{239}$ values for ADs at the same EH are identical to $<$0.1\%.   
Periods of constant positive slope correspond to continuous running and evolution of fuel in the cores, while sharp drops in $F_{239}$ correspond to the shut-down and start-up of a reactor.  
For EH1 (EH2), $\sim$80\% of the antineutrinos originate from the two Daya Bay (four Ling Ao) cores. 
As ADs receive fluxes from multiple cores with differing fuel compositions, variations in the effective fission fractions at an AD are smaller than variations in the fission fractions within a single core.
The relationships between $F_{239}$ and the effective fission fractions of the other fissioning isotopes for the same dataset are shown in the bottom panel of Fig.~\ref{fig:Fracs}.  
The average effective fission fractions $\bar{F}_i$ for $i=(235,238,239,241)$ for the combined EH1 and EH2 ADs were (0.571,0.076,0.299,0.054).

Uncertainties in the input reactor data will result in systematic uncertainties in the measured IBD yields and in the reported $F_{239}$ values.  
The thermal power of each reactor was determined through heat-balance calculations of the reactor cooling water to a precision of 0.5\%, uncorrelated among cores~\cite{bib:cpc_reactor}.  
Dominant uncertainties in this calculation arise from limitations in the accuracy of water flow rate measurements.  
Since these measurement techniques are independent of the core composition, this uncertainty was treated for a single core as fully correlated at all fission fraction values.  
Fission fraction uncertainties of $\delta f_i$/$f_i$=5\% were determined by comparing measurements of isotopic content in spent nuclear fuel to values obtained by the APOLLO2 reactor modeling code~\cite{bib:science2010,bib:cpc_reactor}.  
As these comparisons do not suggest systematic biases in the reported fission fractions for specific burnup ranges, fission fraction uncertainties were treated as fully correlated for all $F_{239}$.  

The fuel evolution analysis is particularly sensitive to detection systematics not fully correlated in time. 
The stability of the ADs' performance in time has been well  demonstrated~\cite{bib:nim_rate,bib:prd_osc}.  
Variations in the detector live time due to periodic calibrations, maintenance, or data quality were corrected for in the analysis with a negligible impact on systematic uncertainties.  
Percent-level yearly time variation in light collection in the ADs has been corrected for in Daya Bay's energy calibration.  
Residual time variations in reconstructed energies of order 0.2\% had negligible impact on the observed rate and spectrum variations described below.  
Time-independent uncertainties in the IBD detection efficiency were also included in the analysis; AD-uncorrelated and AD-correlated efficiency uncertainties are 0.13\% and 1.9\%, respectively~\cite{bib:prd_osc}.

\begin{figure}[htb!pb]
\centering
\includegraphics[width=0.9\linewidth]{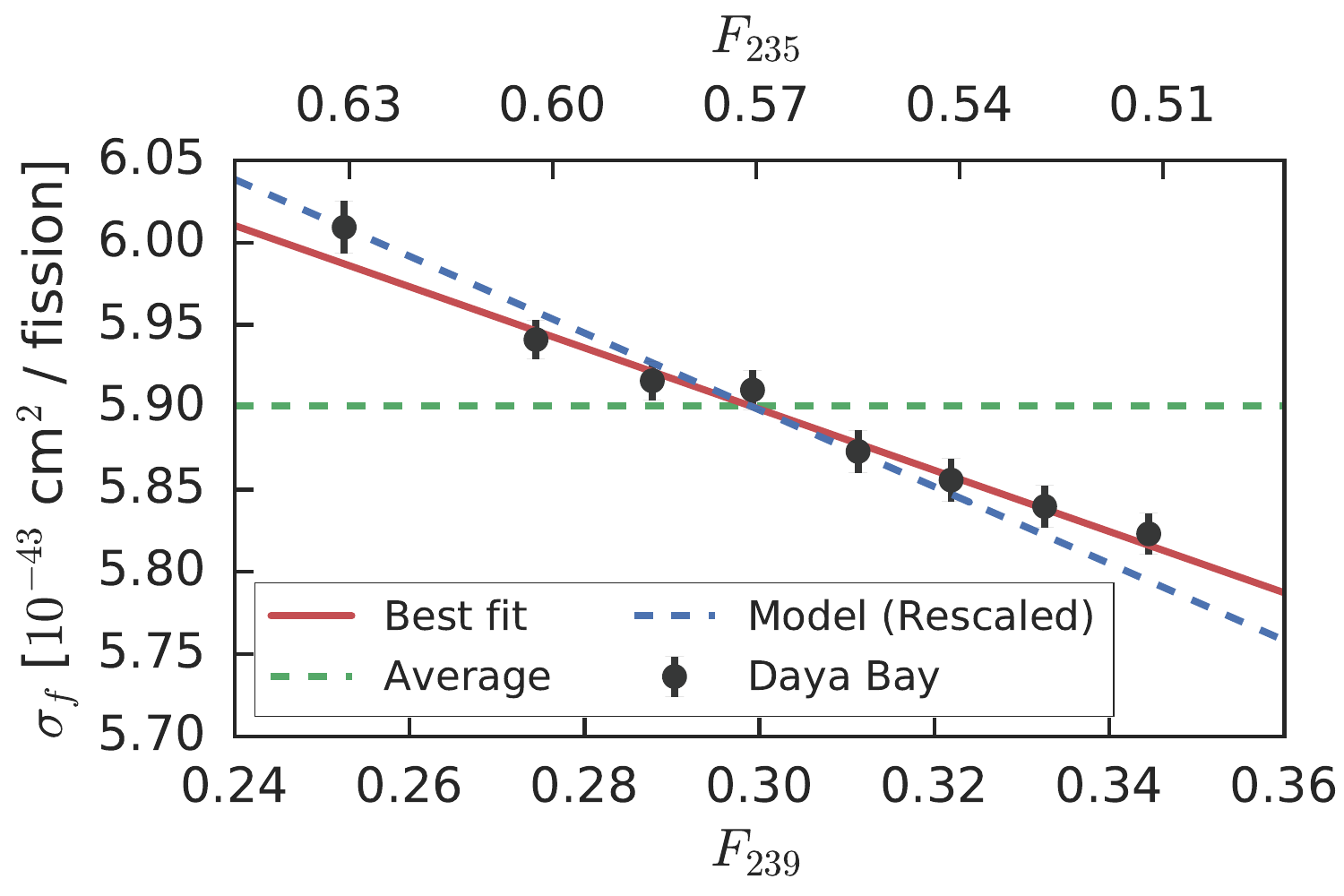}
\caption{IBD yield per fission, $\sigma_f$, versus effective $^{239}$Pu (lower axis) or $^{235}$U (upper axis) fission fraction. 
Yield measurements (black) are pictured with bars representing statistical errors, which lead the uncertainty in the measured evolution, $d\sigma_f/dF_{239}$.  
Constant yield (green line) and variable yield (red line) best fits described in the text are also pictured, as well as predicted yields from the Huber-Mueller model (blue line), scaled to account for the difference in total yield $\bar{\sigma}_f$ between the data and prediction.}
\label{fig:Flux_Change}
\end{figure}

To examine changes in the observed IBD yield and spectrum with reactor fuel evolution, effective fission fractions $F_{239}$ were used to group weekly IBD datasets into eight bins of differing fuel composition, resulting in similar statistics in each bin.  
For the $F_{239}$ bins utilized in this analysis, the effective fission fractions ($F_{235}$, $F_{238}$, $F_{239}$, $F_{241}$) vary within envelopes of width (0.119, 0.001, 0.092, 0.025), as illustrated in Fig.~\ref{fig:Fracs}.  
Each bin's IBD yield per fission, $\sigma_f$ in cm$^2$/fission, was then calculated based on that bin's IBD detection rate~\cite{bib:cpc_reactor}.   
Measured IBD yields~\footnote{Supplementary material included with this Letter provides IBD yield values, covariance matrices, and detailed binning information}, presented in Fig.~\ref{fig:Flux_Change}, show a clear downward trend with increasing $F_{239}$.  

The data were then fit with a linear function describing the IBD yield as a function of $F_{239}$, in terms of the average $^{239}$Pu fission fraction $\overline{F}_{239}$ given above:
\begin{equation}\label{eq:Flux2}
\sigma_f(F_{239}) = \bar{\sigma}_f + \frac{d\sigma_f}{dF_{239}}
(F_{239}-\overline{F}_{239}).
\end{equation}
The fit parameters are the total $F_{239}$-averaged IBD yield $\bar{\sigma}_f$ and the change in yield per unit $^{239}$Pu fission fraction $d\sigma_f/dF_{239}$.  
This fit determines $d\sigma_f/dF_{239}~=~(-1.86~\pm~0.18)\times10^{-43}~\textrm{cm}^2/\textrm{fission}$ with a $\chi^2$/NDF of 3.5/6.
The statistical errors in $\sigma_f$ values are the leading uncertainty in the measurement, with reactor data systematics also providing a non-negligible contribution; errors arising from assuming linear trends in IBD yield with $F_{239}$ (Eq.~\ref{eq:Flux2}) are negligible.  
The fit also provides a total IBD yield $\bar{\sigma}_f$ of (5.90~$\pm$~0.13)~$\times10^{-43}$ cm$^2$/fission with the error dominated by uncertainty in the estimation of the ADs' IBD detection efficiency.
This result was then compared to a constant reactor antineutrino flux model, where $d\sigma_f/dF_{239}$ = 0.  
This model, depicted by the horizontal green line in Fig.~\ref{fig:Flux_Change}, provides a best fit with $\chi^2$/NDF  = 115/7.  
The best-fit $d\sigma_f/dF_{239}$ value is incompatible with this constant flux model at 10 standard deviations ($\sigma$).  
%

Observed IBD yields were compared to those predicted by recent reactor antineutrino models, generated according to Eqs.~\ref{equ_ibd_rate} and~\ref{equ_antinu_prod}.  
Among many available models~\cite{bib:vogel,bib:ILL_1,bib:ILL_2,bib:dwyer}, $^{235}$U, $^{239}$Pu, and $^{241}$Pu antineutrino spectrum per fission  predictions from Huber~\cite{bib:huber} and $^{238}$U predictions from Mueller~\textit{et. al}~\cite{bib:mueller2011} were used to enable a direct comparison to the reactor antineutrino anomaly.  
The predicted total IBD yield $\bar{\sigma}_f$, (6.22~$\pm$~0.14)~$\times10^{-43}$ cm$^2$/fission, differs from the measured $\bar{\sigma}_f$ by 1.7$\sigma$.  
This 5.1\% deficit is consistent with previous measurements reported by Daya Bay~\cite{bib:prl_reactor,bib:cpc_reactor}, as well as with the $\sim$6\% deficit  observed in global fits of past reactor experiments.  
The predicted $d\sigma_f/dF_{239}$ from the Huber-Mueller model, ($-2.46 \pm 0.06) \times 10^{-43} \textrm{cm}^2 / \textrm{fission}$, is represented in Fig.~\ref{fig:Flux_Change} after scaling by the 5.1\% difference in the predicted and measured $\bar{\sigma}_f$ from this analysis.  
This predicted $d\sigma_f/dF_{239}$ differs from the measurement by 3.1$\sigma$, indicating additional tension between the flux measurements and models beyond the established differences in total IBD yield $\bar{\sigma}_f$.  
In particular, it suggests that the fractional difference between the predicted and measured antineutrino fluxes may not be the same for all fission isotopes.  
If the measured fractional yield deficits from all isotopes are equal, the ratio of the slope $d\sigma_f/dF_{239}$ to the total yield $\bar{\sigma}_f$ will be identical for the measurement and prediction.
These ratios, -0.31~$\pm$~0.03 and -0.39~$\pm$~0.01, respectively, are incompatible at 2.6$\sigma$ confidence level.  

The evolution of Daya Bay's IBD yield pictured in Fig.~\ref{fig:Flux_Change} was also used to measure the individual IBD yields of $^{235}$U and $^{239}$Pu.  
For each $F_{239}$ bin $a$ in Fig.~\ref{fig:Flux_Change}, the measured IBD yield can be described as
\begin{equation}\label{eq:Iso1}
\sigma^a_f = \sum_i F^a_i \sigma_i,
\end{equation}
where $F^a_i$ are the effective fission fractions for each isotope, and $\sigma_i$ is the IBD yield from that isotope.
Measurements from all bins can be summarized with the matrix equation
\begin{equation}\label{eq:Iso2}
\boldsymbol{\sigma}_f = F\boldsymbol{\sigma},
\end{equation}
where $\boldsymbol{\sigma}_f$ is an eight-element vector of the measured IBD yields, $\boldsymbol{\sigma}$ is a vector containing the IBD yields of the four fission isotopes, and $F$ is a 8$\times$4 matrix containing fission fractions for the data in each $F_{239}$ bin.  
This matrix equation was used to construct a $\chi^2$ test statistic
\begin{equation}
\label{eq:Iso3}
\chi^2 = (\boldsymbol{\sigma}_f - F\boldsymbol{\sigma})^{\top}
	\textrm{V}^{-1}
    (\boldsymbol{\sigma}_f - F\boldsymbol{\sigma}),
\end{equation}
which allows a scan over the full $\boldsymbol{\sigma}$ parameter space.  
The matrix V is a covariance matrix containing the previously discussed statistical, reactor, and detector uncertainties, and their correlation between measurements $\boldsymbol{\sigma_f}$.

\begin{figure}[htb!pb]
\centering
\includegraphics[width=0.9\linewidth]{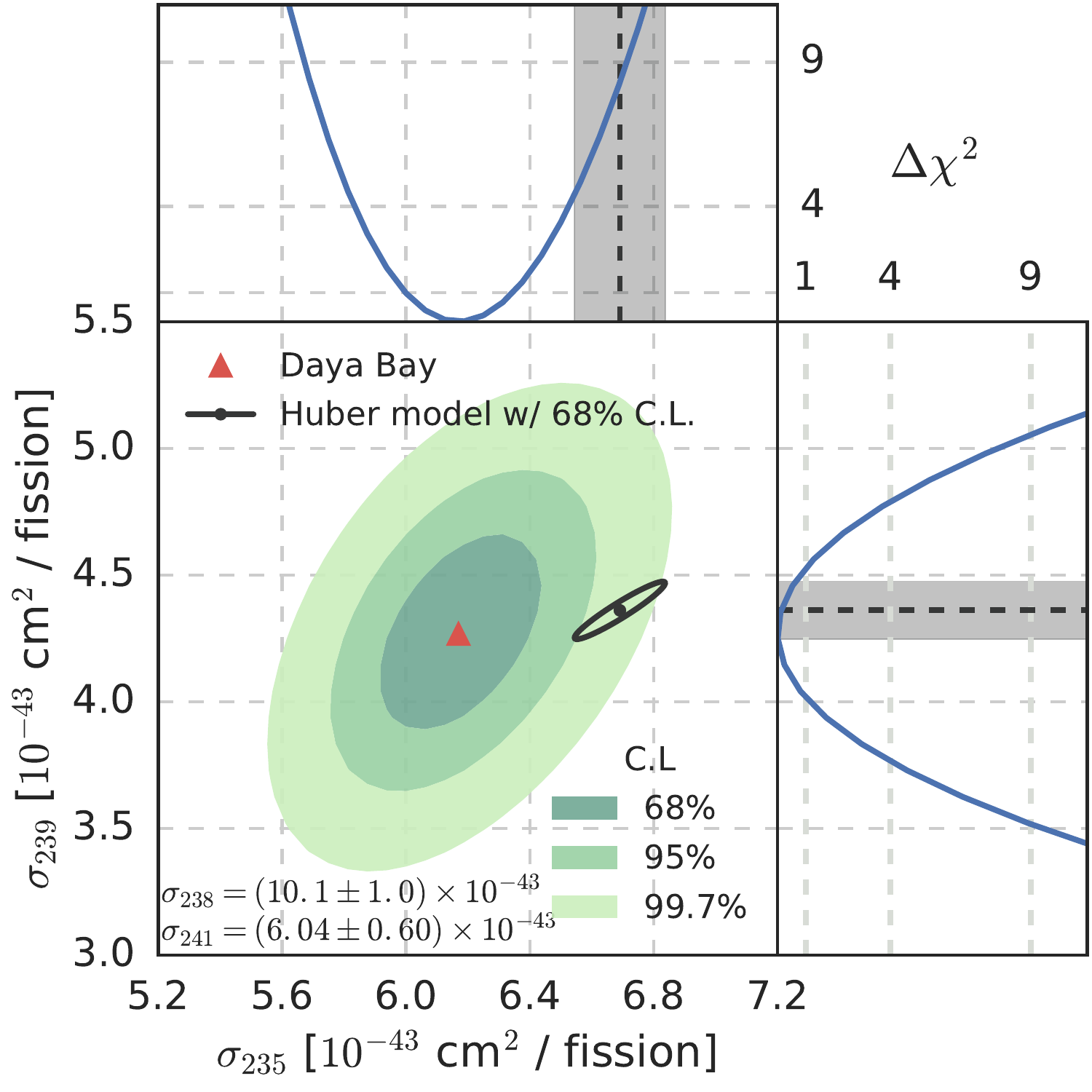}
\caption{Combined measurement of $^{235}$U and $^{239}$Pu IBD yields per fission  $\sigma_{235}$ and $\sigma_{239}$.  The red triangle indicates the best fit $\sigma_{235}$ and $\sigma_{239}$, while green contours indicate two-dimensional 1$\sigma$, 2$\sigma$ and 3$\sigma$ allowed regions.  
Contours utilize theoretically predicted IBD yields for the subdominant isotopes $^{241}$Pu and $^{238}$U as indicated in the lower left panel.  
Predicted values and 1$\sigma$ allowed regions based on the Huber-Mueller model are also shown in black.  
The top and side panels show one-dimensional $\Delta \chi^2$ profiles for $\sigma_{235}$ and $\sigma_{239}$, respectively.
}  
\label{fig:contours}
\end{figure}

In order to break the degeneracy from contributions of the two minor fission isotopes $^{241}$Pu and $^{238}$U, weak constraints were applied to these isotopes' IBD yields.  
This was accomplished in Eq.~\ref{eq:Iso3} by adding terms $(\sigma_i-\hat{\sigma}_i)^2$/$\epsilon_i^2$ for $^{238}$U and $^{241}$Pu, where $\hat{\sigma}_i$ and ${\epsilon}_i$ are theoretically predicted IBD yields and assigned uncertainties, which were treated as fully uncorrelated.
Values for $\hat{\sigma}_i$ were taken from Ref.~\cite{bib:mueller2011} for $^{238}$U (10.1$\times$10$^{-43}$ cm$^2$/fission) and Ref.~\cite{bib:huber} for $^{241}$Pu ( 6.05$\times$10$^{-43}$ cm$^2$/fission).  
Values ${\epsilon}_i$ were set at 10\% of the model-predicted yield, significantly higher than the quoted Huber-Mueller uncertainties, in order to reduce the potential bias to the fit.


The IBD yields from $^{235}$U and $^{239}$Pu, $\sigma_{235}$ and $\sigma_{239}$, were found to be ($6.17 \pm 0.17$) and ($4.27 \pm 0.26$)~$\times 10^{-43}$~cm$^2$/fission, respectively. 
Allowed regions and one-dimensional $\Delta \chi^2$ profiles for $\sigma_{235}$ and $\sigma_{239}$ are shown in Fig.~\ref{fig:contours}.   
The measurement is currently limited in precision by the AD-correlated uncertainty in Daya Bay's detection efficiency, and by the statistical uncertainty in the measurements $\boldsymbol{\sigma_f}$.  
The 10\% uncertainties assigned to $\sigma_{238,241}$ provide a subdominant contribution to the uncertainty in $\sigma_{235}$ and $\sigma_{239}$.  
This $\sigma_{235}$ is 7.8\% lower than the Huber-Mueller model value of ($6.69 \pm 0.15$)~$\times 10^{-43}$~cm$^2$/fission, a difference significantly larger than the 2.7\% measurement uncertainty.  A measured $\sigma_{235}$ yield deficit has also been reported using global fits to antineutrino data from reactors of varying fission fractions~\cite{Giunti}. 
The measured $\sigma_{239}$ value is consistent with the predicted value of ($4.36 \pm 0.11$)~$\times 10^{-43}$~cm$^2$/fission within the 6\% uncertainty of the measurement.  

By applying additional constraints on $\boldsymbol{\sigma_f}$ in Eq.~\ref{eq:Iso3},  these $\sigma_{235}$ and $\sigma_{239}$ results were tested for consistency with hypothetical $\boldsymbol{\sigma_f}$ values representing differing sources of the reactor antineutrino anomaly.  
If the anomaly is produced solely via incorrect predictions of $^{235}$U, the measured $\sigma_{235}$ should deviate from its predicted value while $\sigma_{238,239,241}$ remain at their predicted values; enforcement of this additional constraint in  Eq.~\ref{eq:Iso3} produced a best fit higher by $\Delta \chi^2/$NDF$ = 0.17/1$ (two-sided p-value 0.68).  
A similar test of $^{239}$Pu as the sole source of the anomaly yielded a best-fit value  higher by $\Delta \chi^2/\text{NDF} = 10.0/1$ (p-value 0.00016).  
Requiring all isotopes in Eq.~\ref{eq:Iso3} to exhibit an equal fractional deficit with respect to prediction, the best fit was found to be higher by $\Delta \chi^2/$NDF$ = 7.9/1$ (p-value 0.0049).    
Thus, the hypothesis that $^{235}$U is primarily responsible for the reactor antineutrino anomaly is favored by the Daya Bay data, with the equal deficit and $^{239}$Pu-only deficit hypotheses disfavored at the 2.8$\sigma$ and 3.2$\sigma$ confidence levels, respectively.  

\begin{figure}[htb!pb]
\includegraphics[width=0.9\linewidth]{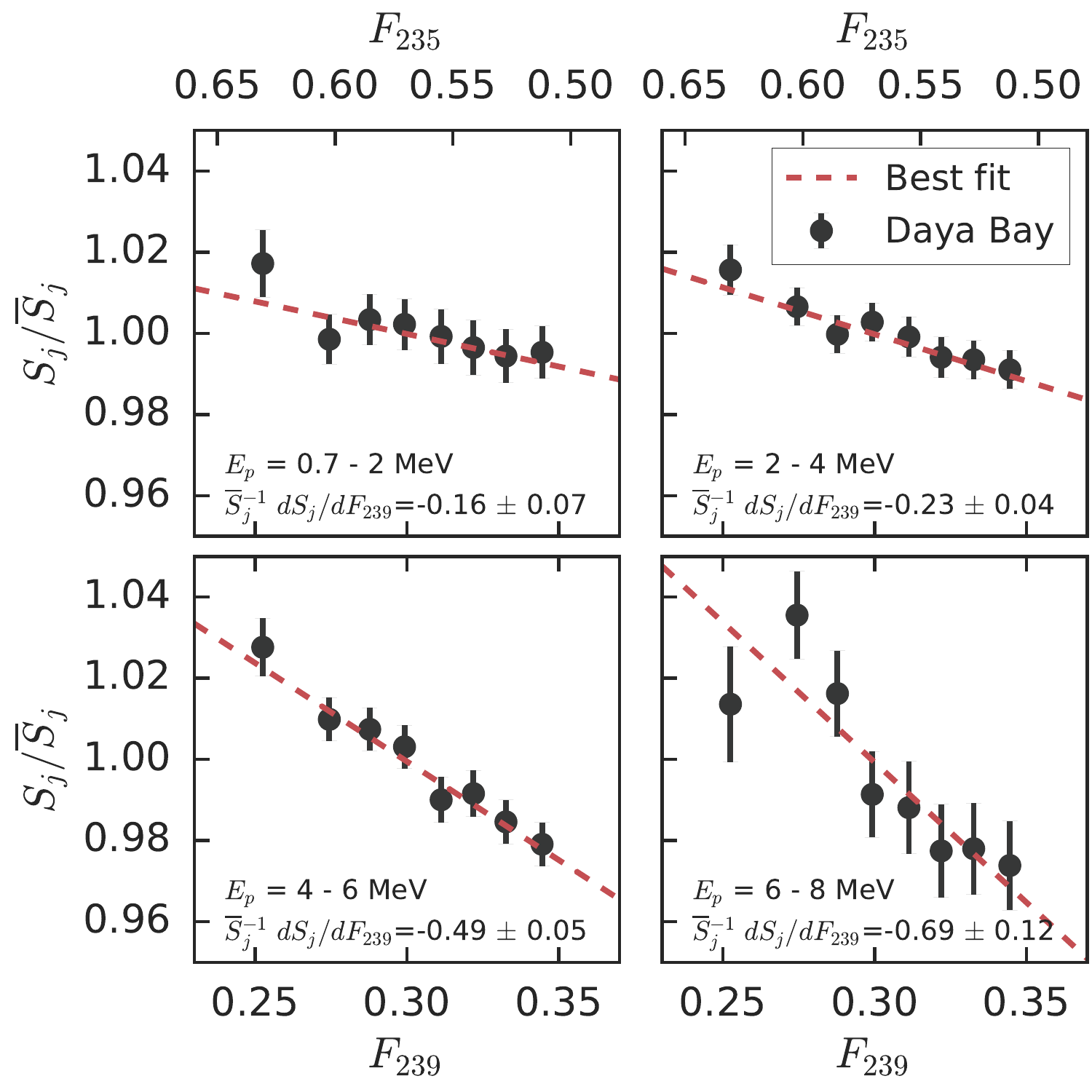}
\caption{Relative IBD yield per fission versus effective $^{239}$Pu (lower axis) or $^{235}$U (upper axis) fission fraction for different prompt energy $E_p$ ranges. The observed slopes $\frac{1}{S}\frac{dS}{dF_{239}}$ are listed in each panel.
}
\label{fig:ChangeWithFrac}
\end{figure}

To investigate changes in the antineutrino spectrum with reactor fuel evolution, observed IBD spectra per fission, $S$, were examined, where $\sigma_f = \sum_j S_j$, the sum of IBD yields in all prompt energy bins.  
For each $F_{239}$ bin depicted in Fig.~\ref{fig:ChangeWithFrac}, the measured $S_j$ values were compared to the $F_{239}$-averaged IBD yield per fission value $\overline{S}_j$.  
The ratio $S_j/\overline{S}_j$ is plotted against $F_{239}$ in Fig.~\ref{fig:ChangeWithFrac} for four different $E_p$ bins.  
The common negative slope in $S_j/\overline{S}_j$ visible in all prompt energy ranges indicates an overall reduction in reactor antineutrino flux with increasing $F_{239}$, as demonstrated in Fig.~\ref{fig:Flux_Change}.
In addition, the trends in $S_j/\overline{S}_j$ with $F_{239}$ in Fig.~\ref{fig:ChangeWithFrac} differ for each energy bin, indicating a change in the spectral shape with fuel evolution.  
In particular, the content of higher-energy bins decreases more rapidly than lower-energy bins as $F_{239}$ increases.  

To quantify the statistical significance of these trends,
a $\chi^2$ fit similar to that of Eq.~\ref{eq:Flux2} was applied to each of the four energy ranges in Fig.~\ref{fig:ChangeWithFrac}:
\begin{equation}\label{eq:Flux3}
S_j(F_{239}) = \overline{S}_j + \frac{dS_j}{dF_{239}} (F_{239} - \overline{F}_{239}).
\end{equation}
If no change in the spectrum shape is observed, $\frac{1}{\overline{S}_j}\frac{dS_j}{dF_{239}}$ values in Fig.~\ref{fig:ChangeWithFrac} should be identical for all energy ranges.
The best-fit $\frac{1}{\overline{S}_j}\frac{dS_j}{dF_{239}}$ value for this scenario is -0.31 $\pm$ 0.03, with a $\chi^2$/NDF of 57.1/27.  
If a change in the spectrum shape is present, each energy range may exhibit an independent $\frac{1}{\overline{S}_j}\frac{dS_j}{dF_{239}}$ value.  
Best-fit $\frac{1}{\overline{S}_j}\frac{dS_j}{dF_{239}}$ values for this scenario, given in the sub-panels in Fig.~\ref{fig:ChangeWithFrac}, produce a $\chi^2$/NDF of 22.6/24.  
The $\Delta \chi^2$/NDF between the best-fit alternative and null hypotheses is 34.5/3, corresponding to the rejection of the hypothesis of no change in the spectral shape at 5.1$\sigma$ significance.   

\begin{figure}[htb!pb]
\centering
\includegraphics[width=0.9\linewidth]{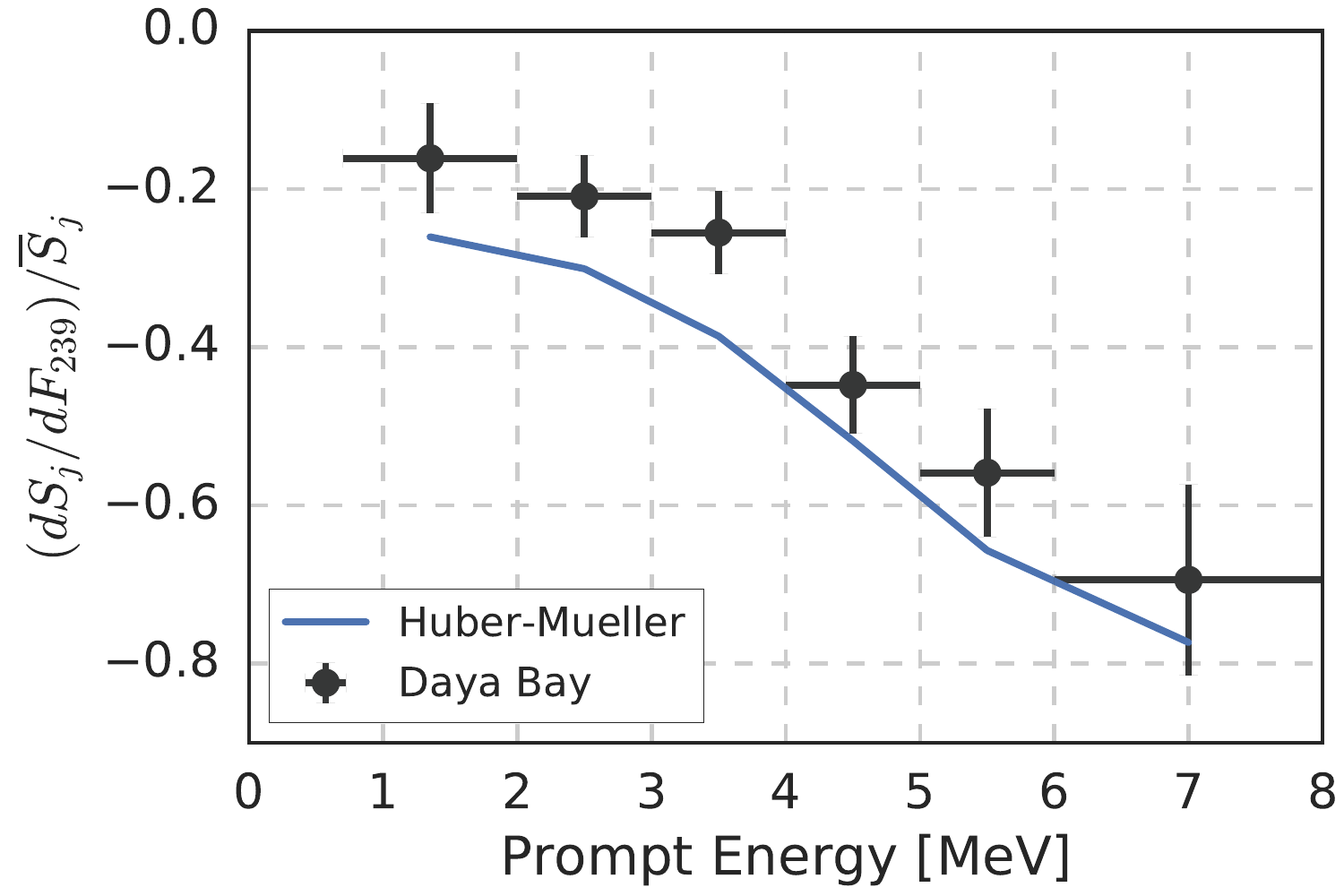}
\caption{Fractional variations in IBD yield $\frac{1}{\overline{S_j}}\frac{dS_j}{dF_{239}}$ for different prompt energy $E_p$ ranges for Daya Bay data and for the Huber-Mueller model.  The variation of the slope $\frac{1}{\overline{S_j}}\frac{dS_j}{dF_{239}}$ with energy, which indicates evolution-dependent changes in the antineutrino spectrum, appears consistent between the data and predictions.}
\label{fig:SpecAsym}
\end{figure}

Measured changes in the IBD spectrum with $F_{239}$ were also compared to that predicted by the Huber-Mueller model.  
To allow direct a comparison to the measured IBD spectrum per fission, antineutrino spectra predicted by the Huber-Mueller model were processed with a detector response matrix to obtain predicted spectra in terms of IBD prompt energy $E_p$~\cite{bib:prd_osc}.  
This comparison is shown in Fig.~\ref{fig:SpecAsym}, where the best-fit slopes in IBD yield per fission $\frac{1}{\overline{S}_j}\frac{dS_j}{dF_{239}}$ are plotted for six prompt energy ranges for the data as well as for the Huber-Mueller model. 

The trend of the measured spectral evolution described by the best-fit $\frac{dS_j}{dF_{239}}$ values is similar to that of the Huber-Mueller model.
This result generally demonstrates the validity of recent theoretical studies describing antineutrino-based monitoring of reactor fissile content~\cite{hubermon1,hubermon2}.  
The data suggest slightly better agreement in $\frac{dS_j}{dF_{239}}$ with the Huber-Mueller model above 4~MeV prompt energy than below, emphasizing the possibility of disagreements in the evolution of both the flux and the spectrum.  
Increased statistics are required in order to investigate the possible isotopic origin of the excess in the observed antineutrino flux from 4-6~MeV prompt energy~\cite{bib:prl_reactor, reno_bump, dc_bump}, a topic discussed recently in the literature~\cite{haser,hayes2,Huber:2016fkt,Huber:2016xis,Giunti}.

In summary, the evolution of Daya Bay's detected IBD yield and energy spectrum has been measured using 2.2~million IBD candidates detected over 1230 days of data taking.  
A total IBD yield $\bar{\sigma}_f$ of (5.90~$\pm$~0.13)~$\times10^{-43}$ cm$^2$/fission was measured with average effective fission fractions $F_{235}$, $F_{238}$, $F_{239}$, and $F_{241}$ of 0.571, 0.076, 0.299, and 0.054, respectively.  
A change in the IBD yield, $d\sigma_f/dF_{239}$, of $(-1.86~\pm~0.18)\times10^{-43}~\textrm{cm}^2/\textrm{fission}$ was observed over a range of effective $^{239}$Pu fission fractions from 0.25 to 0.34.  
These yield measurements were used to calculate IBD yield per fission values of ($6.17 \pm 0.17$) and ($4.27 \pm 0.26$)~$\times 10^{-43}$~cm$^2$/fission for the dominant fission isotopes $^{235}$U and $^{239}$Pu, respectively.  
A change in the IBD energy spectrum with the effective $^{239}$Pu fission fraction was also observed at the 5.1$\sigma$ confidence level.  

These observations were compared to the Huber-Mueller reactor antineutrino model.  
While the measured evolution of the IBD energy spectrum is generally consistent with this model, measured $\bar{\sigma}_f$ and $d\sigma_f/dF_{239}$ values are incompatible with predictions at 1.7$\sigma$ and 3.1$\sigma$ confidence levels.  
These discrepancies indicate issues in modeling the reactor antineutrino flux.  
One can invoke a model including only eV-scale sterile neutrino oscillations to explain the observed deficit in $\bar{\sigma}_f$.  
Such a model requires an equal fractional flux deficit from all fission isotopes and a ratio of $d\sigma_f/dF_{239}$ to $\bar{\sigma}_f$ unchanged from the prediction, which is incompatible with Daya Bay's observation at 2.6$\sigma$.  
A comparison of measured and predicted $^{235}$U and $^{239}$Pu IBD yields instead indicates a preference for an incorrect prediction of the $^{235}$U flux as the primary source of the reactor antineutrino anomaly.  
Improvement in Daya Bay's measurements of $\sigma_{235}$ and $\sigma_{239}$ can be achieved with increased statistics and with a reduction of the AD-correlated IBD detection efficiency systematic uncertainty.  
Future short-baseline experiments at highly enriched uranium reactors~\cite{prospect,stereo,solid} may also provide the capability to probe this apparent overprediction via precise new measurements of the $^{235}$U antineutrino flux.  

Daya Bay is supported in part by the Ministry of Science and Technology of China, the U.S. Department of Energy, the Chinese Academy of Sciences, the CAS Center for Excellence in Particle Physics, the National Natural Science Foundation of China, the Guangdong provincial government, the Shenzhen municipal government, the China General Nuclear Power Group, Key Laboratory of Particle and Radiation Imaging (Tsinghua University), the Ministry of Education, Key Laboratory of Particle Physics and Particle Irradiation (Shandong University), the Ministry of Education, Shanghai Laboratory for Particle Physics and Cosmology, the Research Grants Council of the Hong Kong Special Administrative Region of China, the University Development Fund of The University of Hong Kong, the MOE program for Research of Excellence at National Taiwan University, National Chiao-Tung University, and NSC fund support from Taiwan, the U.S. National Science Foundation, the Alfred~P.~Sloan Foundation, the Ministry of Education, Youth, and Sports of the Czech Republic, the Joint Institute of Nuclear Research in Dubna, Russia, the National Commission of Scientific and Technological Research of Chile, and the Tsinghua University Initiative Scientific Research Program. We acknowledge Yellow River Engineering Consulting Co., Ltd., and China Railway 15th Bureau Group Co., Ltd., for building the underground laboratory. We are grateful for the ongoing cooperation from the China General Nuclear Power Group and China Light and Power Company.

\bibliographystyle{apsrev4-1}
\bibliography{Reactor_8AD_PRD}{}

\end{document}